\documentstyle[epsbox]{mn}
\def\lsim{\mathrel{\mathpalette\Oversim<}}
\def\gsim{\mathrel{\mathpalette\Oversim>}}
\def\Oversim#1#2{\lower0.5ex\vbox{\baselineskip0pt\lineskip0pt%
            \lineskiplimit0pt\ialign{%
          $\mathsurround0pt #1\hfil##\hfil$\crcr#2\crcr\sim\crcr}}}

\newcommand{\apj}{ApJ}
\newcommand{\mnras}{MNRAS}
\newcommand{\aap}{A\&A}

\title{Collapse of Low Mass Clouds in the Presence of UV Radiation Field}
\author[Susa \& Kitayama]
{Hajime~Susa,$^1$\thanks{e-mail:susa@rccp.tsukuba.ac.jp} and Tetsu~Kitayama$^2$\thanks{e-mail:tkita@phys.metro-u.ac.jp}\\
$^1$Center for Computational Physics, University of
  Tsukuba, Tsukuba 305, Japan\\
$^2$Department of Physics, Tokyo Metropolitan University, Hachioji,
Tokyo, Japan }
\date{Accepted 2000 April}
\pubyear{2000}

\begin{document}
\maketitle


\begin{abstract}
The collapse of marginally Jeans unstable primordial gas clouds in the
 presence of 
UV radiation field is discussed. Assuming that the dynamical
collapse proceeds approximately in an isothermal self-similar fashion,
we investigate the thermal evolution of collapsing central core until
H$_2$ cooling dominates photoheating and the temperature drops to
below $10^4$K.  Consequently, the mass of the cooled core is evaluated
as $M_{\rm cool}=3.6\times 10^6 M_\odot
\left(I_{21}/1\right)^{-0.32}$.  This scale depends only on the
incident UV intensity, and provides a lower limit to the mass of
collapsed objects in the UV radiation field.
\end{abstract}

\begin{keywords}
galaxies: formation --- radiative transfer --- --- molecular
processes
\end{keywords}

\section{Introduction}
\label{INTRO}
Formation of primordial objects, such as young galaxies and globular
clusters, in the early universe is a fundamental problem in modern
cosmology.  Rapid progress of observations in the last decade has
pressed theoreticians to construct the physically correct and proper
theory to solve this problem.
Recently, formation of primordial objects has been investigated mainly
in the following two contexts; one is the formation of ``first stars''
or ``first luminous objects''
(e.g. \cite{Tegmark97,Abel98,ON98,NS99}), and the other is that of
``second generation objects'' (e.g. \cite{HRL97,KBS97,ON99}). These
two populations are expected to arise in quite different physical
environments. For instance, the former is likely to be born from the
primordial gas under little influences of the external radiation
field, except for that of the cosmic microwave background radiation
(CMB).  The latter, on the other hand, is largely affected by the
external UV radiation produced by the former. The UV photons not only
ionize and heat up the pregalactic gas clouds, but also dissociate
H$_2$ in the clouds, which is an important coolant in metal-free
gas clouds. In this paper, we pay particular attention to the
formation of the latter population.

At high redshifts, say $z\gg 5$, ionized bubbles around photon
sources, such as young galaxies or AGN's, are still so small that most
of the gas in the universe is not ionized and photoheated. In this
case, the external radiation affects the formation of primordial
objects only via H$_2$ photodissociation \cite{HRL97}.
On the other hand, gas clouds near the ionizing sources, or those at
lower redshifts ($z\lsim 5$) are ionized and photoheated, before they
collapse and cool. Therefore, the formation of population III objects
close to the ionizing sources and the formation of galaxies at low
redshifts are likely to start from hot ionized media.
Kepner, Babul \& Spergel (1997) investigated this problem in the
context of dwarf galaxy formation under the assumption that
pregalactic clouds are in hydrostatic equilibrium. They found that the
gas transit from H$^+$ phase to H and H$_2$ ones, as radiative
cooling proceeds.  Corbelli, Galli \& Palla (1997) also found similar
phenomena in the rotationally supported hydrostatic gaseous disc.
However, gravitational collapse of ionized gas proceeds almost
isothermally \cite{UI84,TW96,KI99,SU00}, and such a collapse becomes
inevitably dynamical \cite{Larson69}.

In this paper, we investigate H$_2$ cooling in the dynamically
collapsing core in marginally Jeans unstable clouds. If these clouds
are self-gravitating, they are likely to collapse almost spherically
with the temperature kept nearly at $\sim 10^4$K, until the external
radiation field is shielded by the clouds themselves. We thus employ
the isothermal Larson-Penston similarity solution \cite{Larson69}
and solve explicitly non-equilibrium chemical reactions and energy
equation to trace the thermal history of a collapsing central
core. Consequently, the mass and the size of the cooled core are assessed,
and their astrophysical implications are discussed.

\newpage
\section{H$_2$ Cooling in the Collapsing Core}
\label{NUMER}
\subsection{Assumptions and Equations}
We consider the collapse of a self-gravitating primordial gas cloud
exposed to the isotropic UV radiation field. If the UV intensity is
strong enough to penetrate through the cloud, the temperature of the
cloud is first kept at $\sim 10^4$K, owing to the balance between
photoheating and radiative cooling.
In this case, gravitational infall proceeds nearly isothermally in a
self-similar fashion \cite{UI84,TW96,KI99,SU00}. The isothermal
collapse of a self-gravitating system always tends to converge to the
Larson-Penston similarity solution \cite{Larson69} because it is a
relatively stable solution (e.g. \cite{HN97}).  We thus assume that
the cloud obeys the Larson-Penston solution, which is characterized by
the central density
\vspace{0.5cm}
\begin{eqnarray}
\rho_{\rm c}(t)=
\frac{\rho_{\rm c}(0)}{\left(t/t_{\rm LP}\left(0\right)-1\right)^2},
\label{eq:LPS}
\end{eqnarray}
and the core radius 
\vspace{0.5cm}
\begin{eqnarray}
r_{\rm c}(t)=\left(t_{\rm LP}\left(0\right) - t \right) c_{\rm s}.
\label{eq:CORE}
\end{eqnarray}
Here $t$ denotes the elapsed time from an initial instant characterized
by $\rho_{\rm c}(0)$. 
$c_{\rm s}$ is the sound speed in the core, 
and $t_{\rm LP}$ is defined as 
$t_{\rm LP}\left(0\right)\equiv \sqrt{1.667/4\pi G \rho_{\rm c}
\left(0\right)}$ 
In order to elucidate the thermal history of nearly isothermally
collapsing core, we employ an one-zone approximation and integrate
numerically the following energy equation: 
\vspace{0.5cm}
\begin{eqnarray}
\frac{d\epsilon_{\rm c}}{dt}&=&\frac{P_{\rm c}}{\rho_{\rm
c}^2}\frac{d\rho_{\rm c}}{dt}-\frac{\Lambda - \Gamma}{\rho_{\rm c}},
\end{eqnarray}
where $P_{\rm c}$ is the central pressure, $\epsilon_{\rm c}$ is the
specific energy per unit mass, and $\Lambda$ and $\Gamma$ denote the
cooling and heating rates, per unit volume, respectively.  The cooling
rate includes H$_2$ cooling by rovibrational transition lines, as well
as H atomic cooling. We also solve time-dependent non-equilibrium
reactions for e, H, H$^+$, H$^+_2$, H$^-$, and H$_2$, to evaluate
$\Lambda$. Unless stated explicitly, reaction rates are taken from the
recent compilation by Galli \& Palla (1998). H$_2$ photodissociation
rate is evaluated using the self-shielding function in Draine \&
Bertordi (1996). H$_2^+$ photodissociation is also taken into account,
based upon the cross section in Stancil (1994). H$^-$ radiative
detachment is assessed using the fitting formula for the cross section
in Tegmark et al.(1997).

The incident UV intensity is set to be $I_\nu^{\rm in} =
I_{21}(\nu/\nu_L)^{-1} 10^{-21} {\rm erg \; s^{-1} cm^{-2} str^{-1}
Hz^{-1}}$, where $\nu_L$ is the frequency at the Lyman limit.  We take
various values for $I_{21}$ in the range $10^{-3}\lsim I_{21} \lsim
10^3$.  The power-law approximation of the UV spectrum adopted
here is admittedly rather crude. In reality, the UV intensity in the
Lyman-Werner bands could be modulated by so-called the ``sawtoothing
effect'' and reduced by a factor of $10-10^{2}$ at $z=15$
\cite{HAR00}. As will be discussed later, however, the two-step
photodissociation process is not important in the present case and the
spectral change in the Lyman-Werner bands will not alter our
results.\footnote{The sawtoothing effect could also suppress
destruction of H$_2^+$ by the UV photons, which could in turn enhance
the H$_2$ abundance. The assumption of a power-law spectrum should
thus provide the limiting case in which H$_2$ formation is maximally
suppressed. In practice, however, this has only minor impacts on our
present results (see also Section 3).}

The photoionization rate is computed taking account of
the absorption of ionizing photons as \cite{TU98,SU00}:
\begin{eqnarray}
k_{\rm ion}&=& n_{\rm HI}\int_{\nu_L}^\infty \int \frac{I_\nu}{h\nu} \sigma_{\nu} d\Omega d\nu \nonumber\\
&\simeq& \frac{I_{\nu_L}^{\rm in}\sigma_{\nu_L}}{h}\int
\frac{1}{3\tau_{\nu_L}(\Omega)^{4/3}}\gamma(4/3,\tau_{\nu_L}\left(\Omega\right))d\Omega \nonumber \\
&\simeq& \frac{4\pi}{3}\frac{I_{\nu_L}^{\rm in}n_{\rm HI}\sigma_{\nu_L}}{h}
\frac{\gamma\left(4/3,\tau_{\nu_L}\right)}{\tau_{\nu_L}^{4/3}}\hspace{0.5cm}
{\rm [\; s^{-1} cm^{-3}]}, 
\label{eq:pion}
\end{eqnarray}
where $\sigma_{\nu}$ is the photoionization cross section which is
proportional to $\nu^{-3}$, $\Omega$ is the solid angle, and
$\gamma(a,b)$ represents the incomplete gamma function. The optical
depth at the Lyman limit is defined as $\tau_{\nu_L}\equiv n_{\rm HI}
r_{\rm c} \sigma_{\nu_L}$, where $r_{\rm c}$ is assessed by equation
(\ref{eq:CORE}) and $n_{\rm HI}$ denotes the central HI number
density. The accreting envelope outside $r_{\rm c}$ only alters the
optical depth $\tau_{\nu_L}$ less than a factor of 2. Similarly, the
photoheating rate is expressed as
\begin{eqnarray}
\Gamma&=& n_{\rm HI}\int_{\nu_L}^\infty \int \frac{I_\nu}{h\nu}
\sigma_{\nu} (h\nu-h\nu_L)d\Omega d\nu \nonumber \\
& \simeq& \frac{4\pi}{3}I_{\nu_L}^{\rm in} n_{\rm HI}\sigma_{\nu_L}\nu_L
\left(\frac{\gamma(1,\tau_{\nu_L})}{\tau_{\nu_L}}
-\frac{\gamma\left(4/3,\tau_{\nu_L}\right)}{\tau_{\nu_L}^{4/3}}\right) \nonumber\\
&~&\hspace{4cm}{\rm [\; erg\; s^{-1} cm^{-3}]}.
\label{eq:pheat}
\end{eqnarray}

The initial central density is taken as twice the self-shielding
critical density $n_{\rm cr}$ derived in Tajiri \& Umemura (1997),
where $n_{\rm cr}$ represents the density above which the HI fraction
exceeds 0.1.  Thus, the numerical integrations of the energy equation
and chemical reactions start when the cloud center begins to be almost
neutral. 
The initial temperature and chemical abundances are
fixed assuming thermal and chemical equilibria.  Even if they are
perturbed, they immediately converge to the equilibrium values.  We
stop the numerical integrations when the temperature of the core drops
to below $5000$K and the assumption of an isothermal cloud breaks
down.

\subsection{Numerical Results}

\begin{figure}
\begin{center}
\psbox[width=150mm,vscale=1.0]{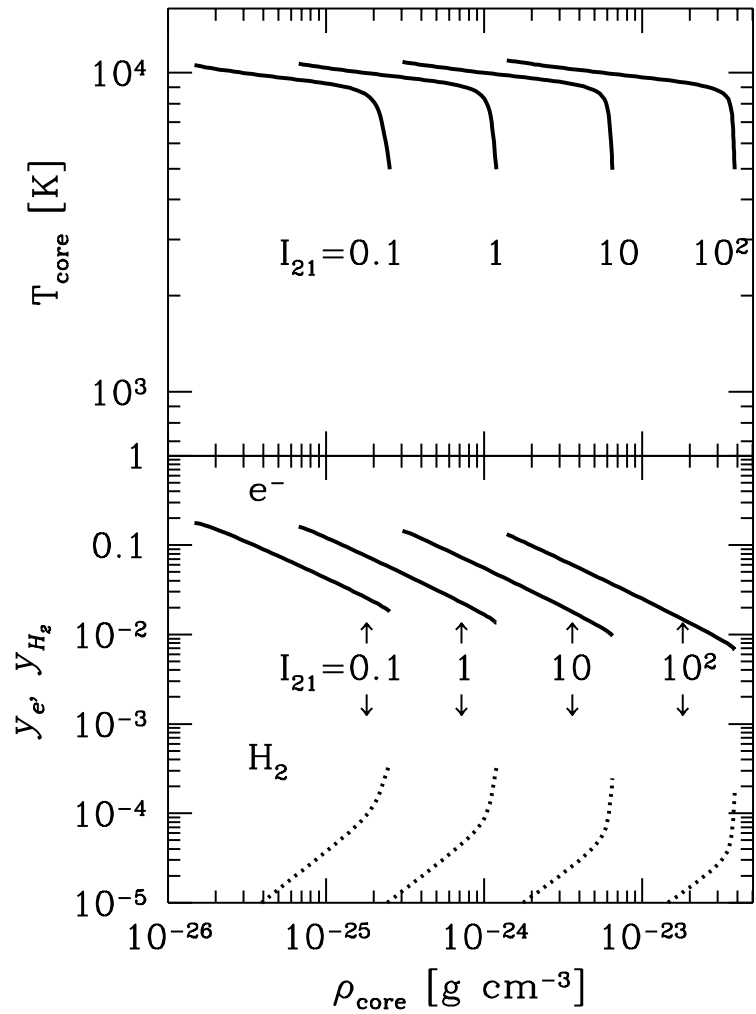}
\end{center}
\caption[f1.eps]{Time evolution of the physical quantities in the
 collapsing core for four different values of the incident UV
 intensity $I_{21}$.  The upper panel shows the evolution of
 temperature, while the lower panel that of electron fraction (solid
 lines) and H$_2$ fraction (dotted lines).}
\label{fig1}
\end{figure}

\begin{figure}
\begin{center}
\psbox[width=150mm,vscale=1.0]{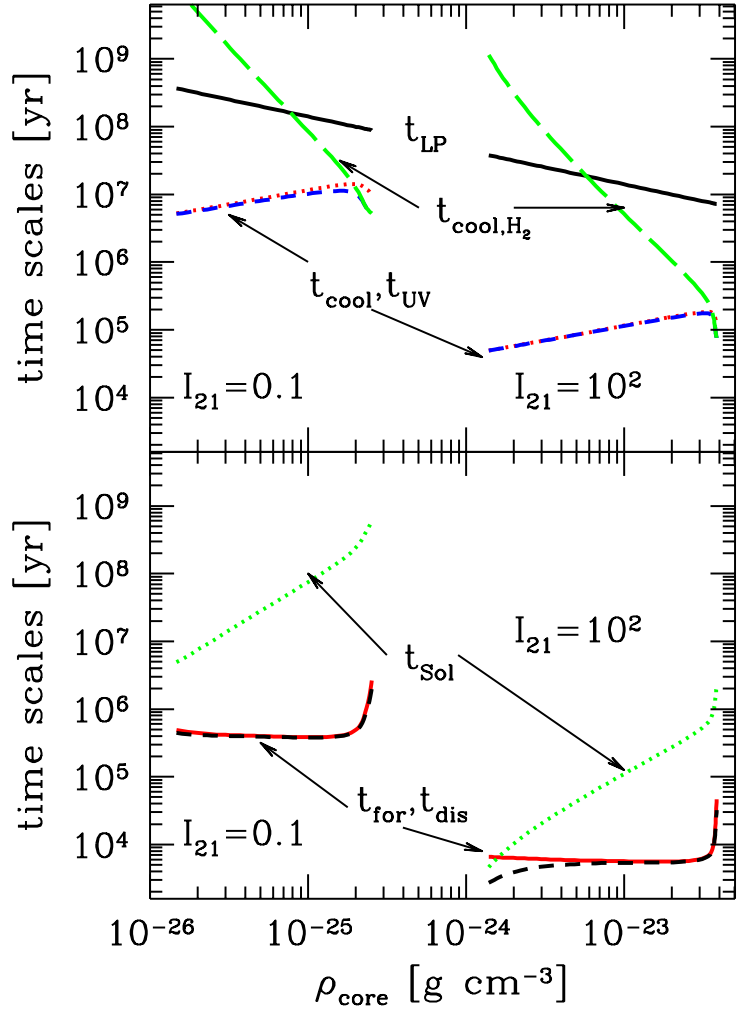}
\end{center}
\caption[f2.eps]{Various time-scales during the evolution for
$I_{21}=0.1$ and $10^2$. The upper panel shows the collapse time of
the Larson-Penston similarity solution (solid line; $t_{\rm LP}$), the
total cooling time (short dased; $t_{\rm cool}$), the H$_2$ radiative
cooling time (long dashed; $t_{\rm cool, H_2}$), and the photoheating
time (dotted; $t_{\rm UV}$). In the lower panel, time-scales related
to H$_2$ formation and dissociation are plotted; the H$_2$
photodissociation time (dotted; $t_{\rm Sol}$), the H$_2$ collisional
dissociation time (solid; $t_{\rm dis}$), and the H$_2$ formation time
(short dashed; $t_{\rm for}$).}
\label{fig2}
\end{figure}

The numerical results are presented in Figs.\ref{fig1} and \ref{fig2}.
In Fig.\ref{fig1}, time evolution of the core temperature (upper
panel) and the chemical compositions (lower panel) are plotted for
four different values of the incident intensity. In every case, the
temperature evolves nearly isothermally after the collapse calculation
starts. This ensures that the density evolution approximately follows
the isothermal Larson-Penston similarity solution. After the
isothermal collapse, core temperature drops dramatically, with the
time-scale much faster than the collapse time, and the loci on the
$\rho_{\rm c}-T_{\rm c}$ plane become nearly vertical. This
corresponds to the transition from the H phase to the H$_2$ one
mentioned in Kepner, Babul and Spergel (1997). The increased cooling
rate by H$_2$ overwhelms the photoheating rate which is reduced due to
strong self-shielding.  In the lower panel of Fig.\ref{fig1},
fractions of electron and H$_2$ are plotted. At $T\gsim 10^4$ K, the
electron fraction is kept at $\sim 10^{-1}$ by ionizing UV photons. At
$T\lsim 10^4$ K, the photons are self-shielded and recombination
proceeds. However, the radiative cooling time becomes shorter than the
recombination time, and the electrons become out of equilibrium.
Significant amount of free electrons thus ``freeze out'' as the system
cools via Ly-$\alpha$ and H$_2$ ro-vibrational transitions.
The ionization degree still remains
at the level of $\sim 10^{-2}$, even at $T\sim 5000$K.  Feeded these
relic electrons, H$_2$ molecules are formed to the level of $\sim
10^{-3}$. Remark that the similar phenomena have been found in the
postshock region of primordial gas \cite{SK87,KS92,Susa98,Nishi98}.

In Fig.\ref{fig2}, various time-scales are plotted for $I_{21}=0.1$
and $10^2$. The upper panel shows time-scales related to the energy
equation, i.e., the collapse time ($t_{\rm LP}$), the cooling time
($t_{\rm cool}$), the H$_2$ cooling time ($t_{\rm cool,H_2}$), and the
photoheating time ($t_{\rm UV}$).  It is clear that temperature is
determined by thermal equilibrium, because the condition $t_{\rm UV}
\simeq t_{\rm cool}$ is well satisfied. The transition from the H
phase to the H$_2$ one takes place as H$_2$ cooling dominates UV
heating. We remark that the collapse time-scale $t_{\rm LP}$ is much
longer than the other thermal time-scales. Therefore, hydrostatic
equilibrium assumption is a poor one owing to the cooling instability.

In the lower panel of Fig.\ref{fig2}, time-scales related to H$_2$
formation/destruction are plotted. Chemical equilibrium is achieved
throughout the evolution because the H$_2$ dissociation time is almost
equal to the H$_2$ formation time ($t_{\rm dis}\simeq t_{\rm
for}$). It is clear that two-step photodissociation (Solomon process;
$t_{\rm Sol}$) is not important during this calculation. Dissociation
of H$_2$ is dominated by a collisional process ( H$_2$ + H$^+$
$\rightarrow$ H$_2^+$ + H) at such high temperature
\cite{CGP97,SU00}.

\section{Properties of Cooled Core}
\label{PCC}
As shown in the previous section, the runaway collapsing core cools
rapidly as soon as H$_2$ cooling becomes effective. The core will
probably cool down to $\sim 100$ K, which is the lowest temperature
achievable by H$_2$ cooling.  The cooled core will then be a
free-falling sphere, since the cooling time-scale is much shorter than
the collapse time (Fig.\ref{fig2}).  Consequently, violent collapse of
the cooled core will lead to subsequent star formation in the cloud
center.

\begin{figure}
\begin{center}
\psbox[width=150mm,vscale=1.0]{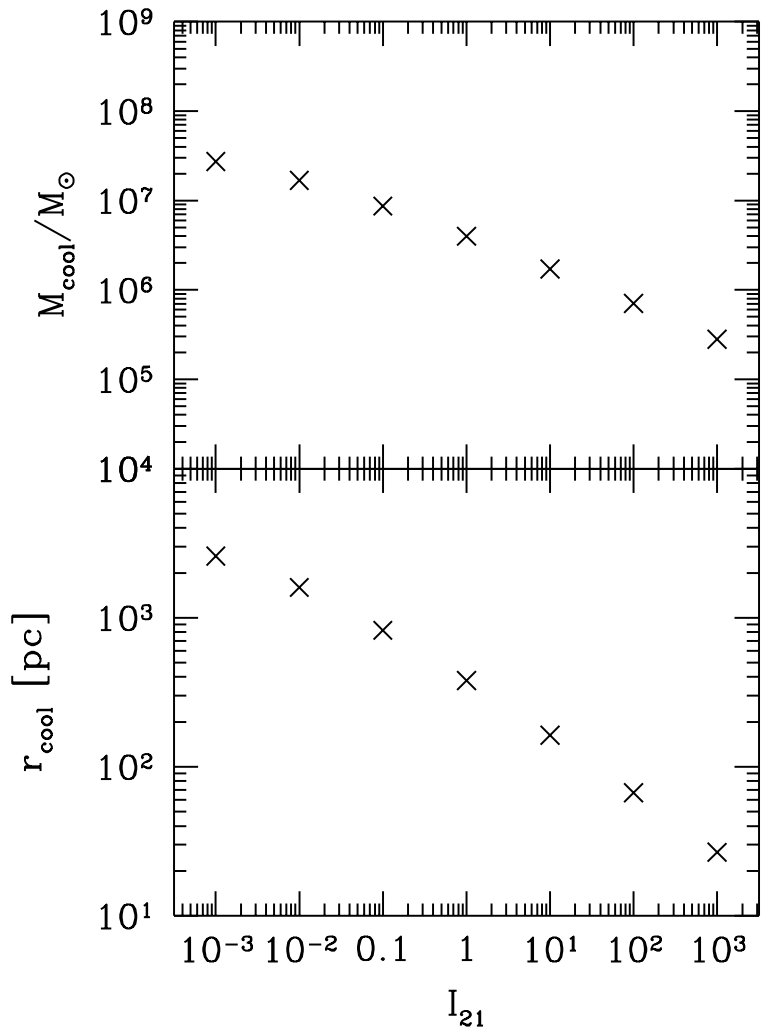}
\end{center}
\caption[f3.eps]{Mass (upper panel) and radius (lower panel) of the
 core at $T_{\rm core} =5000$ K, i.e. at the onset of H$_2$ cooling,
 as a function of the incident UV intensity $I_{21}$.}
\label{fig3}
\end{figure}

In Fig.{\ref{fig3}}, the mass and the radius of the cooled core are
plotted against the incident UV intensity. The upper panel shows the
cooled mass, which is approximately fitted by $M_{\rm cool}=3.6\times
10^6 M_\odot \left(I_{21}/1\right)^{-0.32}$.  
Remark that this scale depends only on the incident intensity
$I_{21}$. This dependence can be understood essentially as
follows. $M_{\rm cool}$ is roughly estimated from three conditions at
the onset of H$_2$ cooling: 1) $M_{\rm cool} \simeq M_{\rm
J}(T=10^4{\rm K})$, 2) $\Gamma \simeq \Lambda_{\rm H_2}$, and 3)
$y_{\rm H_2} \simeq 10^{-4}-10^{-3}$. Condition 1) simply dictates
that the core mass is comparable to the Jeans mass. Condition 2) means
that the thermal equilibrium is achieved until the onset of H$_2$
cooling. Condition 3) is based on the fact that the H$_2$ fraction
(denoted as $y_{\rm H_2}$) at this moment roughly converges to a fixed
level (Fig.\ref{fig1}) \footnote{$y_{\rm H_2}$ at the onset of H$_2$
cooling is basically determined by chemical equilibrium with given
temperature and electron fraction. As the temperature drops to $T\sim
8000$K, the dominant cooling mechanism switches from H Ly-$\alpha$
cooling to H$_2$ cooling. At this temperature, the electron fraction
converges to $10^{-2}-10^{-1}$ regardless of initial conditions
\cite{Susa98,Kita00} and the H$_2$ fraction freezes out nearly at
its equilibrium value.}. These three conditions yield $M_{\rm
cool}\propto I_{21}^{-1/3}$, which is slightly steeper than our
numerical results. This difference is caused by two reasons. The
first one is the destruction of H$_2^+$ and H$^-$ by the UV flux which
are neglected in the above simplified estimation. For large $I_{21}$,
the fractions of H$_2^+$ and H$^-$ are reduced, and the amount of
H$_2$ produced and H$_2$ cooling rate both become smaller. In order to
achieve the balance between cooling rate and photoheating rate, the
system has to collapse even further. As a result, the cooled mass
becomes slightly smaller than the analytic estimation and the $M_{\rm
cool}- I_{21}$ correlation becomes steeper for large $I_{21}$.  The
other reason is the relatively short free-fall time compared to the
cooling time at lower density. For low $I_{21}$, H$_2$ cooling starts
to operate at low density (Fig.\ref{fig1}) when the difference between
the free-fall time and the cooling time scale is small. This makes
gentler the drop of the temperature as a function of density at the
onset of H$_2$ cooling (Fig.\ref{fig1}, upper panel).  Consequently,
the cooled mass becomes larger and the $M_{\rm cool}- I_{21}$
correlation becomes shallower for small $I_{21}$.

In the limit of very strong UV field, H$_2$ formation is completely
suppressed as H$^-$ and H$_2^+$ are both destroyed by the UV
photons. We find that this happens at the threshold intensity of
$I_{21}\simeq 3.6\times 10^4$ in the case of $\alpha=1$.

The radius of the cooled core is also plotted for various $I_{21}$ in
the lower panel.  It also has the same dependence on $I_{21}$ as
$M_{\rm cool}$, $r_{\rm cool}\propto I_{21}^{-0.32}$. Note that the
absolute values of cooling radius shown in Fig.\ref{fig3} are larger
than the Jeans length of the cooled objects, since they are assessed
just at the onset of H$_2$ cooling. In reality, the temperature drops
rapidly from $\sim 10^4$K to $\sim 100$K due to H$_2$ cooling.
Therefore, the corresponding Jeans length should be reduced by a
factor $\sim 10^2$, provided the cooled core mass is constant during
the free-fall collapse. This Jeans length will provide the actual size
of the cooled object.

For $1\lsim I_{21} \lsim 10^3$, the cooled mass is $\sim 10^5 - 10^7
M_\odot$. Such high intensity is realized near the luminous Population
III objects, such as young galaxies or quasars. For instance,
$I_{21}=10^2$ roughly corresponds to the intensity at $10$kpc away
from the center of active galactic nuclei whose luminosity is $10^{44}
{\rm erg ~s^{-1}}$. In this case, the cooled mass in Fig.\ref{fig3}
should provide a lower limit to the mass of the cooled objects around
or inside the young galaxies/quasars. The mass ($\sim 10^6 M_\odot$),
compactness ($\sim$ a few pc, 100 times smaller than $r_{\rm cool}$ in
Fig.\ref{fig3}) and location ($\sim 10$ kpc) of the cooled core
roughly account for those of globular clusters \cite{BM98}. This
result implies that the globular clusters might be formed in the
runaway collapsing core around luminous host galaxies.

For $10^{-3} \lsim I_{21}\lsim 1$, the cooled mass $M_{\rm cool}$ is
$\sim 10^7 - 10^8 M_\odot$, and cooling radius $r_{\rm cool} \lsim
1{\rm kpc}$. These intensities correspond to the UV background
radiation field inferred from so-called the proximity effect of
Ly$\alpha$ forests \cite{BDO88,GCD96}. The actual mass of the cooled
gas is determined by some mechanism (SN, radiation feedback) which
halts the mass accretion from the envelope onto the cooled core.
In this sense, these mass scales provide a lower limit to the mass of
cooled objects under the UV background radiation field. 
HST \cite{Pasca96} and SUBARU \cite{Yamada99}
in fact found the ``building blocks'' which are very compact ($\lsim
1{\rm kpc}$), although the mass is still unsettled. The
creation mechanism of low mass compact objects discussed in this 
paper might be able to explain the formation of such objects.
Finally, let us remark on influences of dissipationless dark matter on
the present results especially in case we apply them to the formation of
dwarf galaxies. If the dark matter dominates the gravity of the
collapsing core, the collapse will not follow the Larson-Penston
similarity solution. The dark matter distribution at the center of a
pregalactic cloud is still highly uncertain, but the cooled mass in
Fig.{\ref{fig3}} is likely to be reduced by the dark matter gravity.
This point will be discussed in more detail in our future publications
\cite{Kita00}.

\section*{Acknowledgments}

 We thank the referee, Tom Abel, for helpful comments,
Masayuki Umemura and Taishi Nakamoto for continuous encouragement, and
Ryoichi Nishi and Yukiko Tajiri for useful remarks. The analysis of
this paper has been made with computational facilities at the Center
for Computational Physics in University of Tsukuba. This work is
supported in part by Research Fellowships of the Japan Society for the
Promotion of Science for Young Scientists, No. 2370 (HS) and 7202
(TK).


\newpage

\newpage
\onecolumn
\end{document}